\PassOptionsToPackage{inline}{enumitem}
\documentclass{ewic}

\usepackage[hyphens]{url}
\usepackage[table,x11names]{xcolor} 
\usepackage{graphicx}
\usepackage{tabularx}
\usepackage[utf8]{inputenc}
\usepackage[english]{babel}
\usepackage{blindtext}
\usepackage{comment}
\usepackage{csquotes}
\usepackage{subcaption} 
\usepackage{natbib}
\usepackage{enumitem} 
\usepackage{balance} 

\graphicspath{ {./images/} } 

\newcommand{\ie}{{\it i.e., }}
\newcommand{\eg}{{\it e.g., }}

\begin{document}

\runningheads{Carr, Chowdhury, Wani, Chiasson}{Understanding Human and Interface Design Factors in Canadian Cybercrime Reporting}

\conference{}

\title{Understanding Human and Interface Design Factors in Canadian Cybercrime Reporting}

\authorone{Charlotte Carr, Ananta Chowdhury, Asra Sakeen Wani, Sonia Chiasson  \\
Carleton University, Canada\\
\email{CharlotteCarr@cmail.carleton.ca}; \{AnantaChowdhury, AsraSakeenWani, SoniaChiasson\}@cunet.carleton.ca}

\begin{abstract}
Cybercrime affects a majority of Canadians, yet most incidents go unreported. We conducted two studies to examine the factors influencing cybercrime reporting and the role of interface design in victims' reporting experiences. Our survey provides individual-level insights into the persistent gap in cybercrime reporting in Canada, showing how perceived incident severity and personal characteristics shape reporting behaviour. Our usability study compared reporting with an AI chatbot to an online form; chatbots facilitated more complete reports and led to higher user satisfaction, highlighting how interface design impacts reporting outcomes. 

\end{abstract}

\keywords{Cybercrime reporting, User interface, User study}

\maketitle

\section{Introduction}

In 2022, 70\% of Canadians experienced a cybersecurity incident ~\citep{statcan2023cybercrime}, whereas only 4.6\% of these crimes were reported to a government authority~\citep{madeinca2024cybercrime}. In 2024, Canadians lost over \$638 million to fraud, with just 5-10\% of cases reported~\citep{angus2026Crime}.   
This gap between cybercrime victimization and reporting may reflect how individuals perceive cyber incidents as well as how reporting systems support victims. Prior research has examined cybercrime reporting by individuals in the USA, UK, and Europe~\citep{chen2026harm}, as well as within organizational settings (e.g., employee reporting)~\citep{burda2025organization}, with limited research in the Canadian context.

We examine cybercrime reporting in Canada from two complementary perspectives: (1) human factors influencing reporting decisions, and (2) interface factors shaping reporting experiences. To 
reflect practical challenges encountered by cybercrime victims and law enforcement in Canada, we collaborated with the 
National Cybercrime Coordination Centre (NC3)\footnote{https://rcmp.ca/en/federal-policing/cybercrime/national-cybercrime-coordination-centre}. Following NC3's definition, we define \textbf{cybercrime} as any crime where technology plays a central role, either as the target (e.g., data breaches), or as the means of committing the crime (e.g., identity theft), including cyber-enabled fraud (e.g., phishing, romance scams)~\citep{werksmans2023dichotomy}. We define reporting as formally submitting a cybercrime incident to an official authority, such as a government reporting system or law enforcement cybercrime unit.
Moreover, while prior studies often focus on specific types of cybercrime (e.g., phishing)~\citep{pilavakis2023phishing}, we holistically examine reporting behaviour across five common and emerging incident types identified by the NC3, enabling a more comprehensive understanding of reporting decisions.

Our research questions are: \textbf{(1)} \textit{How do factors such as individual characteristics and perceived severity of cyber incidents influence victims’ cybercrime reporting behaviour?} \textbf{(2)} \textit{How do different reporting interfaces shape users’ reporting experiences, perceptions, and willingness to report cyber incidents?} We conducted two related studies. First, an online survey with 204 participants examined the human factors potentially shaping reporting decisions. 
Second, a usability study with 24 participants compared two reporting interfaces: a conventional online form and a conversational AI chatbot, to see how they shaped reporting experiences. This sequential design enabled us to move from broad behavioural insights to a more applied, interface-focused analysis. Our Research Ethics Board reviewed both studies. Our contributions include empirical insights into: 
1) Why there is a persistent gap between cybercrime prevalence and reporting, 2) The influence of perceived incident severity and personal characteristics on reporting behaviour, 3) How interface design affects report completeness and user satisfaction, and 4) Recommendations for designing cybercrime reporting systems.

\section{Related Work}
\subsection{Cybercrime Reporting and Barriers}

Cybercrime 
reporting to government authorities remains very low in Canada~\citep{madeinca2024cybercrime} and across nations, as also reflected in the review of UK cybercrime victimization by \cite{sikra2023uk}. Many countries have established dedicated reporting platforms, such as Canada's Anti-Fraud Centre (CAFC)\footnote{https://antifraudcentre-centreantifraude.ca/index-eng.htm} and UK Action Fraud\footnote{https://www.reportfraud.police.uk/}.
Yet, these systems face persistent challenges that contribute to under-reporting. 
For instance, \cite{sikra2024scottish} identified systemic barriers in Scottish reporting, including victims’ confusion about which agency to contact and a lack of feedback after reporting. These infrastructural challenges highlight the need to examine not only whether people report, but how reporting systems can better support them.

Beyond infrastructure issues, victims face numerous barriers to reporting. Victims often lack clarity about what constitutes reportable cybercrime, struggle to identify appropriate reporting channels, perceive reporting as complex, and doubt it will lead to meaningful outcomes~\citep{bidgoli2016end}. Being redirected between agencies further discourages reporting~\citep{cross2018victims}. Social stigma compounds these procedural barriers; \eg fraud victims face stereotypes of being ``greedy or gullible,'' discouraging formal reporting~\citep{cross2015no}.

Underreporting is also shaped by how incidents are recognized and interpreted. Many victims fail to identify their experience as a crime~\citep{bidgoli2016cybercrimes} or misclassify what occurred~\citep{breen2022large}. 
\cite{10.1145/3415231} describes a three-stage process where experts identify phishing: sensemaking, suspicion, and investigation. Only after this interpretive work can reporting occur. 

When reporting occurs, motivations are often prosocial; victims seek justice and aim to prevent harm to others~\citep{cross2018victims, van2024cybercrime}
Recent work on phishing reporting confirms and extends these motivations. \cite{burda2025organization} found that employees report primarily to protect colleagues and their organization, while \cite{pilavakis2023phishing} documented users framing reporting as a civic duty. \cite{chen2024employees} demonstrated that training interventions can increase reporting by cultivating both detection skills and support-seeking behaviours. 

\subsection{Technology-Mediated Reporting}
Technology-mediated reporting research shows that interface design shapes both report quality and victims’ experiences. \cite{bidgoli2019report} demonstrated that minimalist interfaces with clear definitions, simple questions, and undo capabilities improve usability and support accurate triage. Participants preferred web forms over phone reporting for non-phishing incidents because forms were self-paced and reduced social discomfort. 

Specific interaction patterns affect report quality and users' emotional burden. \cite{de2023reporting} tested three reporting styles across platforms.
Interfaces offering specific categories produced the most effective reports, while free-text fields enabled detail but increased emotional distress, and general menus often caused confusion. Further, tools prompting tailored follow-up questions can improve report completeness, but may reduce accuracy~\citep{iriberri2010internet}. 
These findings suggest trade-offs between expressiveness, structure, and user burden.

Conversational AI introduces new possibilities for reporting~\citep{bradford2025whom},
while also raising tensions around trust and persistent algorithm aversion. At the same time, chatbots that combine authoritative structure with warmth elicit more accurate reports than those that lack these qualities~\citep{vitro2025chatbot}. People tend to disclose highly sensitive information more readily to computers than to humans due to reduced fear of judgment~\citep{ho2018psychological}. Both empathy cues and active listening increase depth of disclosure~\citep{10.1145/3313831.3376175}. Language formality, interaction modality, and long-term memory also shape disclosure and privacy perceptions~\citep{papneja2025self,10.1145/3613904.3642420}.
This suggests conversational interfaces can support disclosure, but must be carefully designed.
Yet most prior work examines interfaces in isolation from the human and social factors that shape reporting decisions, largely focusing on phishing or workplace security incidents. Direct comparisons between conversational and form-based reporting for broader cybercrimes remain limited. Our work addresses this 
by examining how human factors and interface modality shape cybercrime reporting experiences.

\section{Survey Study}
This study investigated factors influencing cybercrime reporting behaviour in Canada, considering both lived experience and hypothetical cyber incidents. 
We focused on two factors: perceived incident severity and individual characteristics. Severity was measured across three dimensions: \textit{financial loss} (money lost due to the incident), \textit{time} (time and effort required to address the incident), and \textit{emotional impact} (stress or worry caused by the incident). The survey had a total of 136 Likert-scale and multiple-choice items; it employed display logic to present only relevant questions. We refined the survey through multiple pilot tests to ensure clarity and reasonable completion time (approx. 16 minutes). 

Our survey (Appendix~\ref{app.survey})
included five main components. (1) \textbf{Victimization Experience}
captured direct and indirect exposure to cyber incidents, along with reporting behaviour and reasons for not reporting.
(2) \textbf{Impact Assessment} asked participants to assess their most significant incident's severity, impact, and reporting motivations.
(3) To enable consistent assessment of incidents across all participants regardless of personal experiences,  \textbf{Hypothetical Scenarios} presented 11 scenarios (Appendix~\ref{app.survey}) developed in collaboration with the NC3.
(4) \textbf{Psychological Scales} included the New General Self-Efficacy Scale (NGSE)~\citep{chen2001validation} to examine how confidence in navigating formal processes relates to reporting motivations, and the Auckland Individualism-Collectivism Scale (AICS)~\citep{shulruf2007development} to explore whether orientation toward collective welfare influences reporting behaviour~\citep{van2024cybercrime,cross2018victims}. 
(5) \textbf{Demographic Information} included age, gender, race, household income, household size, technical proficiency, and language comfort.

\subsection{Survey Participants}
Participants were recruited in January 2025 through Prolific\footnote{\url{https://www.prolific.com}}; the survey was hosted on Qualtrics\footnote{\url{https://www.qualtrics.com}}, and participants received £2.75 ($\approx$\$5 CAD) upon completion. Eligible participants resided in Canada, were over 18 years old, and could complete the survey in English. Prior cybercrime experience was not required. A total of 214 participants started the survey; 9 incomplete and one low-quality responses were discarded, leaving 204 valid responses. The sample included 50\% men, 47\% women, 2\% non-binary individuals, and 1\% undisclosed.  
Age was evenly distributed across four brackets (18--24, 25--34, 35--49, and 50--64). 
55\% held a 4-year degree or higher, 14\% a 2-year college degree, 29\% were high school graduates, and 1\% reported less than a high school education. Most participants self-rated their computer and internet skills as advanced (44\%) or very advanced (26\%), and (29\%) as intermediate.

\subsection{Survey Results}
Our analysis used 
descriptive statistics, regression modelling, and correlational analysis. Analyses were conducted separately for real-life incidents and hypothetical scenarios. 
All analyses used a significance threshold of $p < 0.05$.

\subsubsection{Cybercrime Experiences and Reporting}
\begin{figure*}[tb]
  \centering
  \begin{minipage}[t]{0.48\textwidth}
    \centering
    \includegraphics[width=\linewidth]{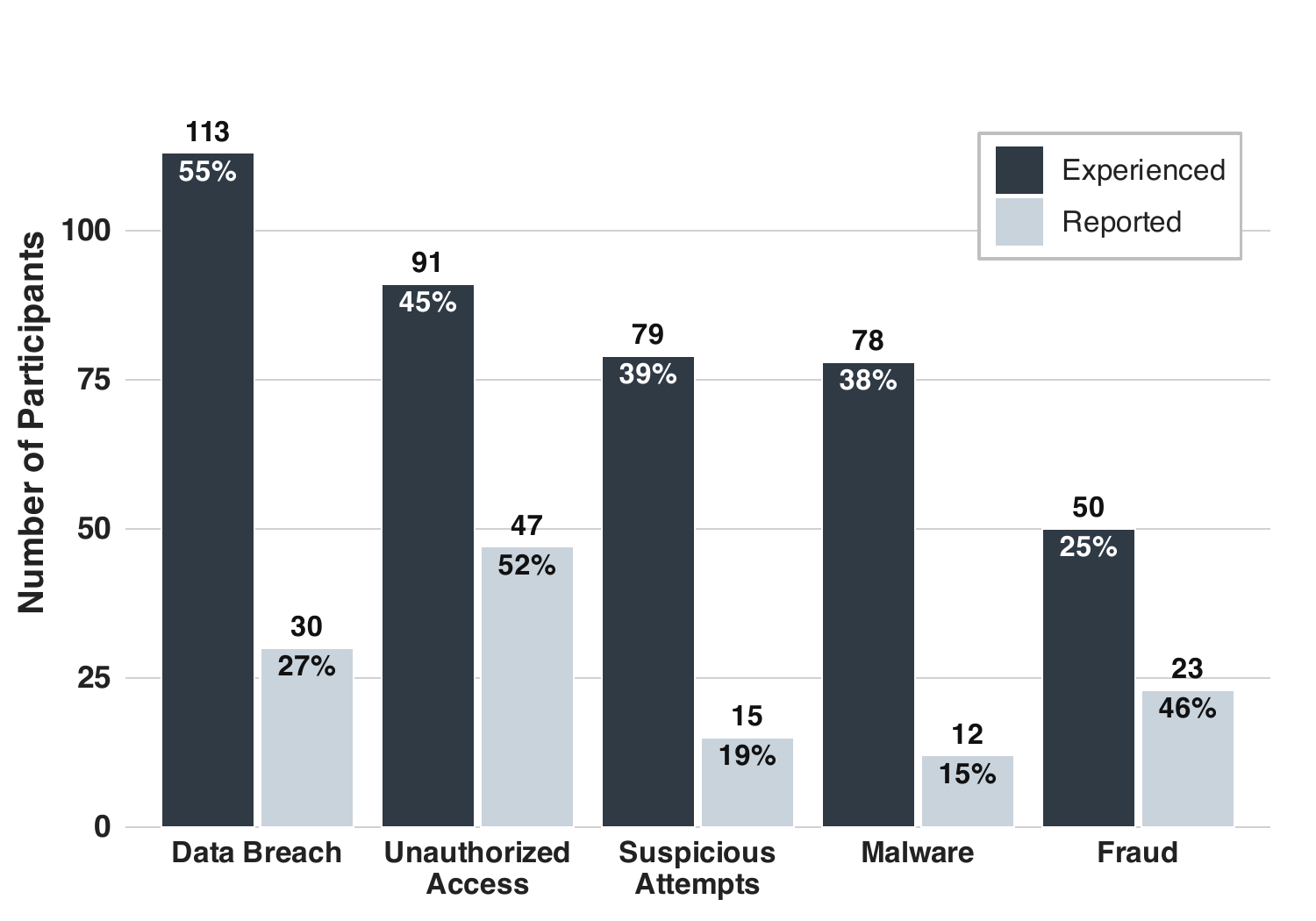}
    \caption[Participants' Reporting Incidents]{Number and percentage of participants experiencing and reporting each type of cyber incident}
    \label{fig:cci}
  \end{minipage}
  \hfill
  \begin{minipage}[t]{0.48\textwidth}
    \centering
    \includegraphics[width=\linewidth]{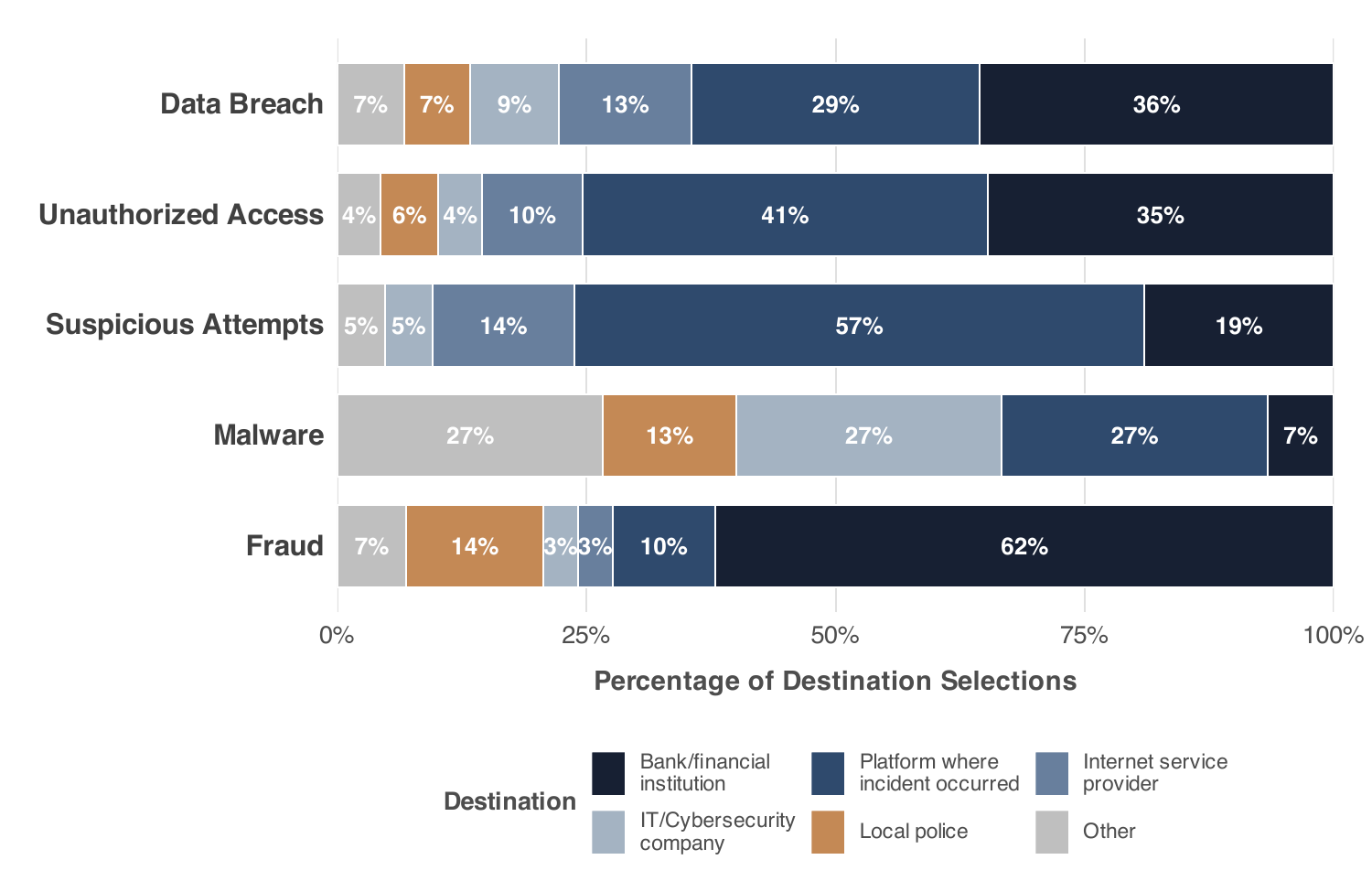}
    \caption{Percentage of participants reporting destinations for each type of cyber incident}
    \label{fig:des}
  \end{minipage}
\end{figure*}

While only 50\% initially identified as cybercrime victims, 91\% reported experiencing at least one cyber incident when presented with specific examples, suggesting that many individuals do not recognize certain incidents as cybercrime. 
Most participants (64\%) experienced multiple cyber incidents, while only 9\% experienced none.
The overall reporting rate was 31\% (127 out of 411), indicating most cyber incidents go unreported even when individuals recognize victimization. Moreover, participants' reporting behaviour varied by incident type (see Figure~\ref{fig:cci}). Data breaches were the most commonly experienced ($n=113$), but only 27\% were reported, whereas fraud ($n=50$, 46\% reported) and unauthorized account access ($n=91$, 52\% reported) were reported more often despite being less prevalent. 
Incident types may differ in expectations, impacts, perceived responsibilities, or reporting pathways, reinforcing the need to consider incident-specific dynamics when designing reporting systems.

Participants' reporting destinations varied by incident type, with reports often directed to organizations involved, such as service providers, affected companies, or financial institutions. For example, fraud incidents were most commonly reported to financial institutions (62\%), and less frequently to law enforcement (14\%) (Figure~\ref{fig:des}). Across all incident types, dedicated cybercrime reporting portals such as the Canadian Anti-Fraud Centre were rarely used, suggesting limited awareness of or barriers to specialized reporting services.

The most common reasons for \textbf{not reporting} were perceiving the incident as not serious enough (24\%), not knowing where to report (22\%), and believing reporting would not help (20\%). Interestingly, (19\%) said they handled the issue themselves, suggesting that participants may see reporting as something that helps them personally, rather than something done for the broader community. Although less common (3\%), embarrassment was also cited as a barrier. These factors suggest that even when participants recognize an incident as a cybercrime, practical barriers and perceptions of resolution can make formal reporting seem unnecessary or difficult.

\begin{figure*}[tb]
\begin{subfigure}{0.5\textwidth}
      \vskip\baselineskip
      \centering
        \includegraphics[width=\linewidth]{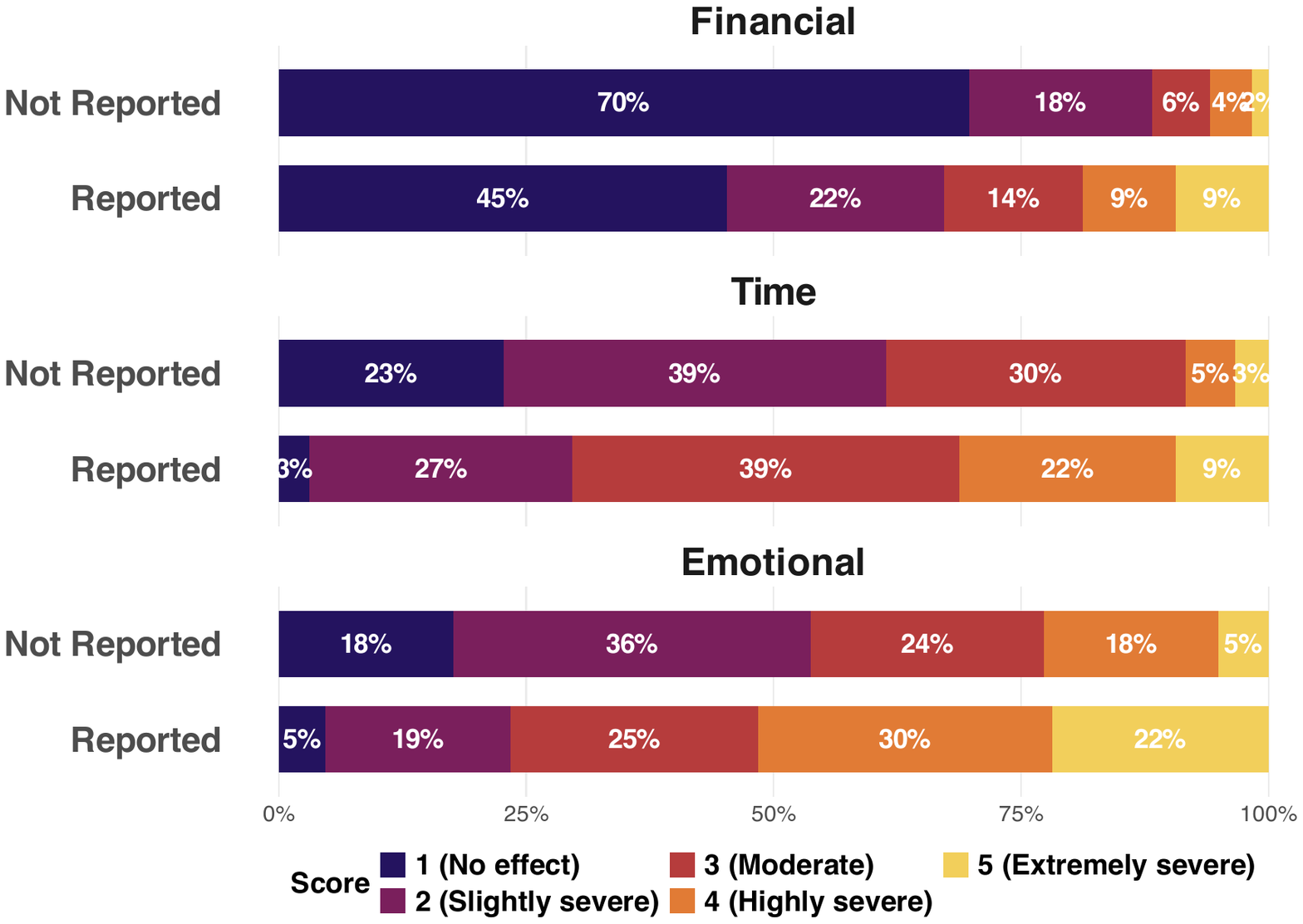}
        \caption{The perceived \textit{financial}, \textit{time}, and \textit{emotional} impact of experienced cyber incidents, organized by whether the participant had reported the event (Reported: n=64, Not reported: n=119)}
        \label{fig:actual_severity}
    \end{subfigure}
    \hfill
    \begin{subfigure}{0.5\textwidth}
    \centering
        \includegraphics[width=\linewidth]{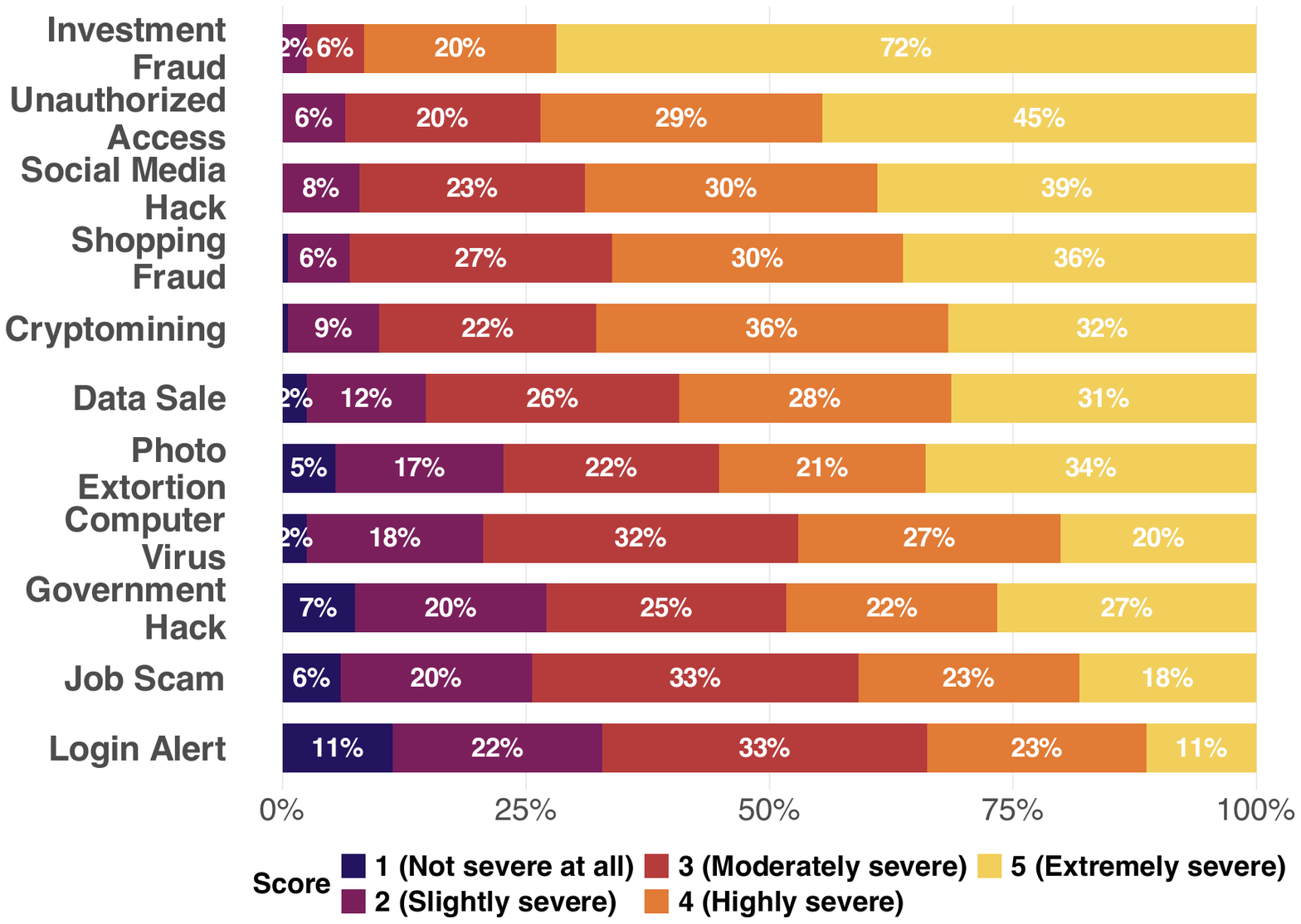}
        \caption{The perceived overall severity of the 11 hypothetical scenarios}
        \label{fig:hypothetical_severity}
    \end{subfigure}
    \caption{Perceived severity of cyber incidents}
    \label{fig:impact}
\end{figure*}

\subsubsection{Incident Severity Influence on Reporting}
Figure~\ref{fig:impact} summarizes the perceived severity of cyber incidents. When considering solely their most significant experienced incidents, participants reported the emotional and time impacts to be more severe. For the hypothetical scenarios, most participants indicated that such an incident would be of at least moderate severity. Participants noted that investment fraud would be the most severe incident, whereas login alerts were perceived as least severe.   

 Logistic regression predicts the outcome of a binary outcome variable based on a predictor variable(s). We used it to determine whether perceived incident severity predicts reporting behaviour, with the overall severity score as the predictor and reporting status as the binary outcome variable.
The severity score was calculated by averaging each participant's ratings of the financial, time, and emotional impacts of their most personally significant incidents. The results showed a significant positive relationship ($\beta = 0.950$, $p < 0.0001$), where each one-point increase in severity was associated with a 2.59-fold increase in the odds of reporting, indicating severity as a strong predictor of reporting behaviour.

Responses to the hypothetical scenarios were measured with Likert scales, so we used Kendall's tau correlation analysis rather than logistic regression.
We measured the correlation between participants' perceived severity Likert response to their likelihood to report Likert response. We found a significant positive correlation across all scenarios ($\tau = 0.781$, $p < 0.001$), indicating that scenarios rated as more severe were also more likely to be reported. This positive association aligns with the pattern that we observed in participants' real-world experiences.

We further examined whether the financial, time, and emotional impact dimensions predicted reporting of experienced incidents. A combined logistic regression model showed that only time was a significant predictor ($\beta = 0.488$, $p = 0.0253$). Financial and emotional impacts were not significant when dimensions were combined.  
These findings suggest that immediacy and effort required to address an incident may influence reporting intentions more strongly than the size of financial loss.

\subsubsection{Personal Characteristics and Reporting}
\textbf{Demographics.}
Using a logistic regression model, we examined how age, gender, technical skill, education, race, and language comfort predict reporting behaviour. After applying stepwise model selection and comparing AIC values, the best-fitting model ($AIC = 379.89$) included age, technical skill, gender, and English comfort.
Technical skill significantly predicted reporting ($p = 0.01$), with each increase in skill level associated with 49\% higher odds of reporting. Gender was also significant, as women were nearly twice as likely to report as men ($p = 0.009$). 
Age was not significant ($p = 0.17$).

\textbf{Psychological Traits.} 
We examined whether self-efficacy (NGSE score) and orientation toward collective welfare (AICS score) influence reporting. Logistic regression showed no relationship between self-efficacy and reporting, suggesting that confidence in one's abilities does not predict whether someone reports cybercrime. 
In contrast, individualism was significantly associated with reporting, where each one-point increase in individualism score increased the odds of reporting by 78\% ($p = 0.02$). This suggests that individuals emphasizing personal rights and individual action are more likely to report cybercrime. 
Cyber incidents involve genuine consequences and personal stakes that could push individuals with individualistic traits to report.
Collectivism showed no significant relationship. 
Kendall's tau showed no statistically significant correlations between hypothetical reporting intentions and either self-efficacy or individualism-collectivism, suggesting that real cyber incidents may activate these traits differently than hypothetical scenarios.

\section{Usability Study}
While the survey study helped us understand \textit{why} people report cybercrime, this usability study investigates how the design of reporting tools shapes the reporting process. Our survey findings revealed that 
the time and effort required to address incidents were more strongly associated with reporting than financial loss, highlighting the importance of effective reporting tools.
To examine these dynamics in a task-based setting, we implemented two prototypes: a conventional online form based on the existing system used by Canada's NC3 and a comparable conversational AI chatbot. We evaluated them using a within-subjects design. Each participant completed a simulated reporting task for each interface, based on two fictional cybercrime scenarios: 1) a non-delivery scam from a fraudulent retail website and 2) a phone scam involving a caller impersonating a Canada Revenue Agency (CRA) representative requesting gift cards. Hypothetical scenarios, developed with the NC3 to ensure realism, enabled all participants to evaluate standardized incidents while avoiding variability from prior personal experience. To control for ordering effects, scenario and interface order were counterbalanced using a $2 \times 2$ Latin Square.

\subsection{Implementing Reporting Interfaces}
To minimize branding effects and colour associations, both interfaces used neutral grey and black tone and displayed the same shield emoji to symbolize protection. Both prototypes were intended to collect the same information.
\begin{figure}[tb]
\centering
\begin{subfigure}{0.4\textwidth}
      \vskip\baselineskip
        \includegraphics[width=.9\linewidth]{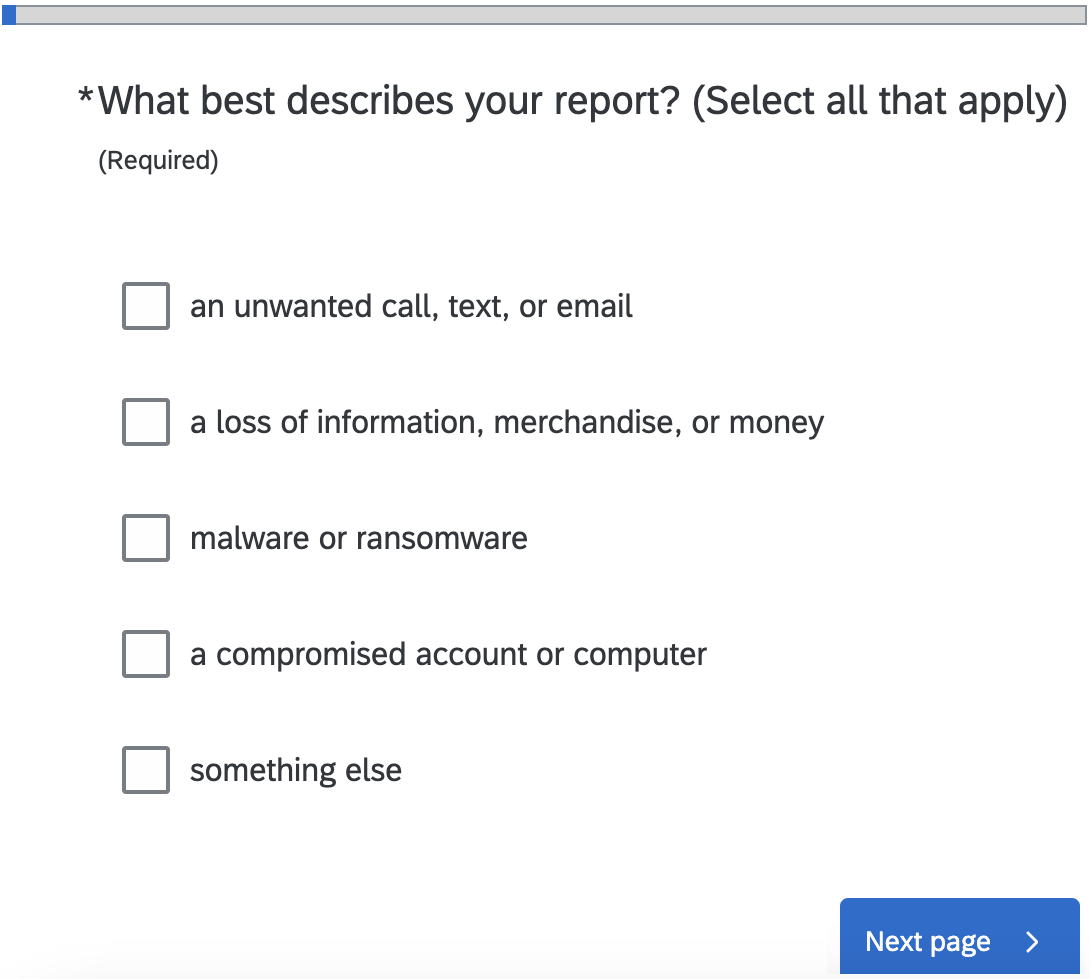}
        \caption{Form interface }
        \label{fig:form}
    \end{subfigure}

    \begin{subfigure}{0.4\textwidth}
        \includegraphics[width=\linewidth]{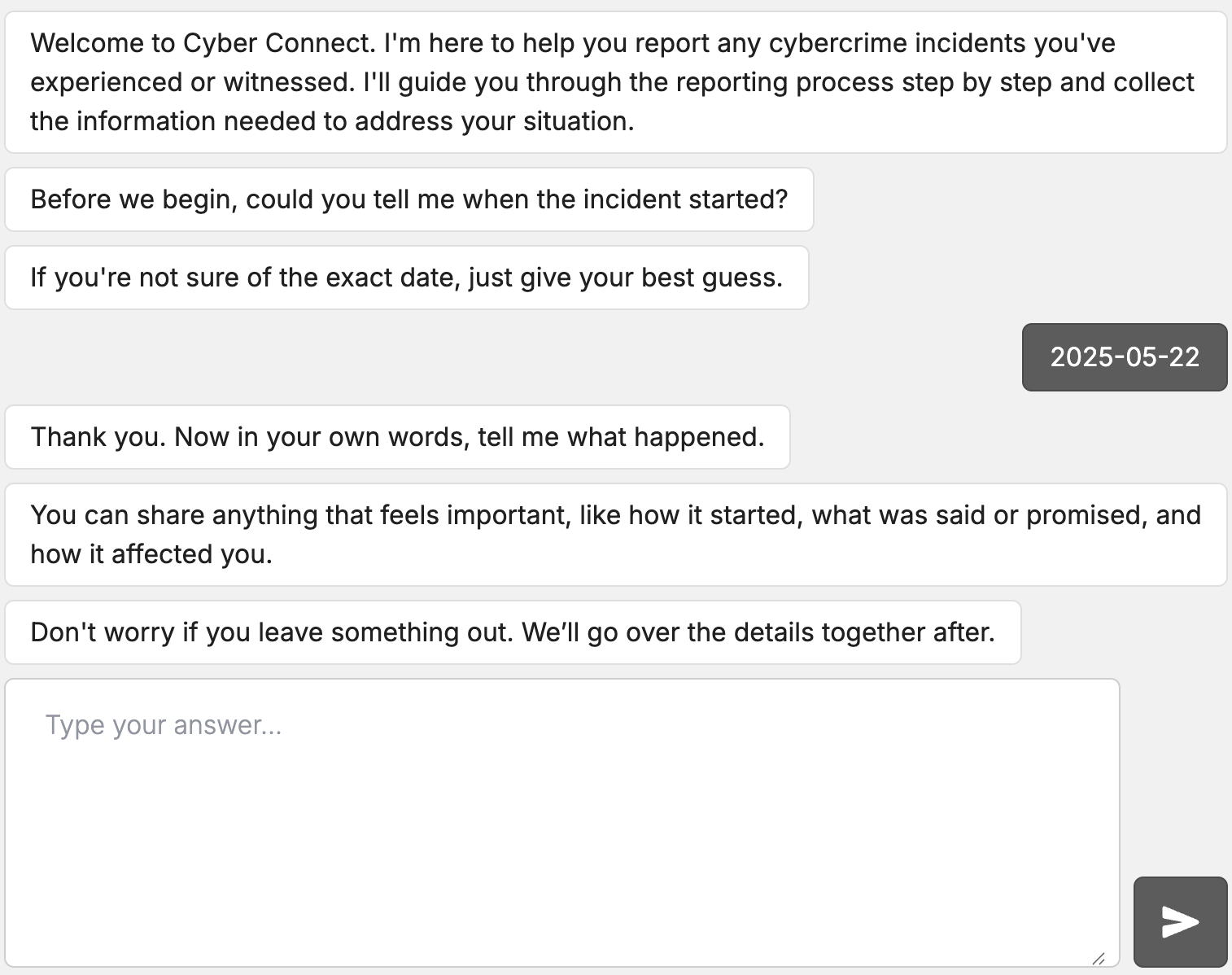}
        \caption{Chatbot interface}
        \label{fig:bot}
    \end{subfigure}
    \caption{Reporting interfaces for the usability study}
\end{figure}

\textbf{Form-Based Interface.}
The form interface was developed in Qualtrics (Figure~\ref{fig:form}) 
following NC3's existing form as closely as possible. Sections outside the scope of this study (\eg business-related incidents, users' contact information) were excluded. The form began with multiple-choice questions and ended with an open-text description of the incident. Mirroring the original form, our interface included the supportive static safety tips presented near relevant questions (\eg advising not to share personal information with strangers), and included options to upload supporting evidence. The form provided a summary of responses upon completion of the report. 
Some features of the original platform could not be replicated in Qualtrics, such as the ability to edit responses before submission and the generation of an official report number. We acknowledge these differences as limitations of the form interface used in our study.

\textbf{AI Chatbot Interface.}
Our chatbot interface (Figure~\ref{fig:bot}) was built using Typebot\footnote{https://typebot.io/} and GPT-4.1 via OpenAI's API\footnote{https://openai.com/index/gpt-4-1/}, enabling a conversational reporting experience. It was designed to reflect NC3's best practices for communication style and accessibility practices. It communicated at a Grade 6 reading level and used a professional and empathetic tone without being informal. The chatbot first asked participants to describe the incident in their own words and then used targeted follow-up questions to collect all required information. We configured the AI model to act as a cybercrime reporting assistant, and provided it with a detailed list of required reporting fields. It was instructed to ask one question at a time, avoid repetition, express empathy appropriately for users' situations, and tailor context-aware follow-up questions based on participants' responses. The system iteratively identified missing information, prompted participants to upload supporting evidence, and generated a structured summary of their report. 

\subsection{Usability Study Tasks}
Sessions were conducted remotely via Zoom. 
Participants were first asked about prior experiences with cyber incidents, their reporting decisions, and the reasons behind those decisions. They were then directed to a test environment, which included a scenario description and reporting interface, and instructed to complete each reporting task as if the scenario had happened to them. Participants shared their screens and were encouraged to think-aloud. After each task, participants completed a short questionnaire assessing satisfaction and perceived ease of use. Next, they answered a post-study questionnaire comparing the two interfaces across several dimensions, including usability, trust, comfort, confidence, and perceived completeness, along with demographic questions. Sessions concluded with a brief semi-structured interview about their preferences, challenges, and overall impressions. This feedback helped clarify and add depth to the survey and reporting data. Study materials are available in Appendix~\ref{app.usability}. 

\subsection{Usability Study Participants}
Twenty-six participants took part in the study.  
Two participants were excluded due to poor data quality, leaving 24 completed sessions (average completion time: 39 minutes). Participants were recruited through Prolific\footnote{https://www.prolific.com}. Eligible participants resided in Canada, were over 18 years old, could participate in English, and were willing to keep their camera on throughout the session. Prior cybercrime experience was not required. Participants received £11~($\approx$ \$20 CAD). Participants ranged in age from 18--60 years (mean = 36, median = 32). The sample included 11 men (46\%) and 13 women (54\%). 71\% participants held a 4-year degree or higher, 17\% were high school graduates, and 13\% two-year diploma holders. Technical proficiency was also high, with most participants rating themselves as advanced (54\%) or very advanced (29\%), 13\% as intermediate, and 4\% as basic.

\subsection{Usability Study Results}
Our quantitative analyses assessed interface usability, reporting completeness, and user preferences. Since participants used both tools, most quantitative comparisons were conducted using the Wilcoxon signed-rank test, appropriate for paired ordinal data and small sample sizes. All analyses used a significance threshold of $p < 0.05$. Qualitative data from think-aloud transcripts and interviews were analyzed inductively to identify common experiences, interface challenges, and overall perceptions of the tools.

\subsubsection{User Experience} 
\begin{figure}[tb]
\raggedright
\includegraphics[width=\columnwidth]{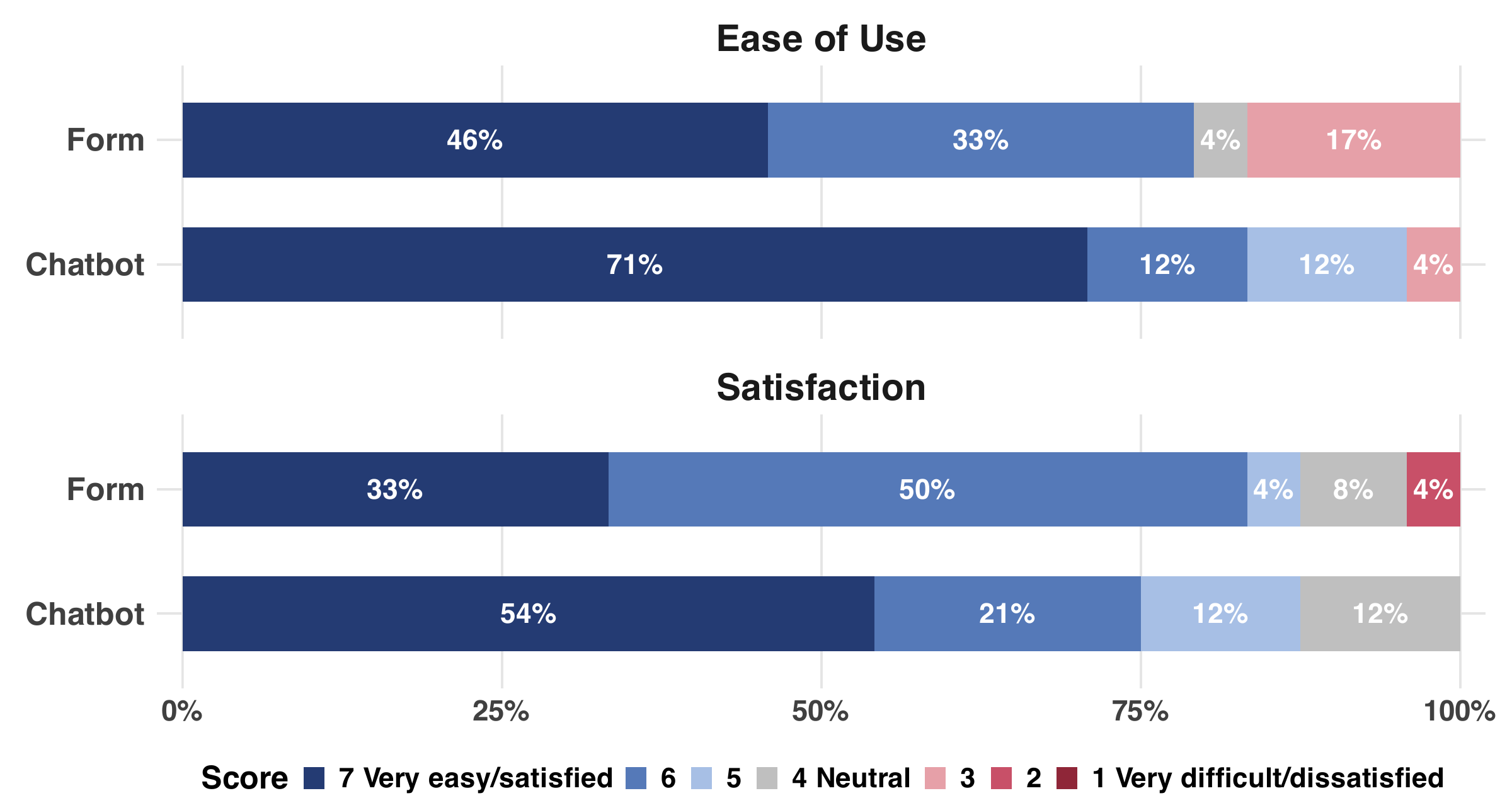}
\caption{Ease of use and satisfaction scores for chatbot and form interfaces}
\label{fig:use}
\end{figure}

Participants rated both interfaces positively (Figure~\ref{fig:use}). The mean ease of use score for the chatbot was 6.46 out of 7 ($\sigma=1.02$), compared to 5.88 ($\sigma=1.48$) for the form. Satisfaction ratings were similarly high, with a mean of 6.17 out of 7 ($\sigma=1.09$) for the chatbot and 5.96 ($\sigma=1.20$) for the form. Wilcoxon signed-rank tests were not statistically significant.

Participants' explanations point to why the chatbot felt easier, even when overall ratings were high for both tools. Many described the chatbot's flow as more supportive and natural. P2 
described it as \textit{``much easier and more comfortable because [...] it basically guided you step-by-step.''} 
In contrast, participants who favoured the form emphasized its directness and control. P11 described it as \textit{``very clear, straightforward, just inputting all the information''}.

\subsubsection{Report Quality}
To assess report completeness, we developed a rubric of incident details per scenario. Each item was scored as 0 (not mentioned), 1 (partially included), or 2 (fully included), depending on how thoroughly the participant addressed that detail and the total was then converted to a percentage. The chatbot reports were significantly more complete than those submitted via the form, with a mean completeness score of 84\% ($\sigma=8\%$), compared to 75\% ($\sigma=15\%$) for the form. A paired t-test indicated that this difference was statistically significant ($t(23)=2.34$, $p=0.028$), with a moderate effect size ($d=0.48$). 

Participant reflections clarify why the chatbot captured more complete details. P2 described the chatbot as \textit{``a guide to [...] which details I should include,''} adding that it made the experience \textit{``much less stressful.''} P12 similarly emphasized that it explicitly asked for \textit{``order numbers, email addresses, and phone numbers,''} and was \textit{``definitely more specific in terms of what information it needed than the form.''} P23 further described it as \textit{``very thorough in asking me almost questions like a police detective would,''} adding, \textit{``I wouldn't have expected to be asked those questions, and I didn't think about adding that detail before being asked.''} However, participants also noted some inconsistencies in the chatbot's prompting. P16 stated, \textit{`` I could have given the confirmation code... and [it] never asked me for that,'' } suggesting that gaps in follow-up can still limit completeness even in a guided interface. P21 noted that the chatbot asked for an order number provided earlier in the conversation: \textit{\enquote{I feel like it should have been able to pick that up [...] instead of having me answer that information again}}.

Perceptions of the form were also mixed. P13 compared it to a multiple-choice test and noted that it \textit{``might've reminded me of certain ideas that were not immediately at the top of my mind,''} suggesting that predefined categories supported recall. Others felt the form offered less guidance about what mattered. P16 described it as \textit{``totally up to your [own] devices what you need to report,''} and wished it were \textit{``kind of like the chatbot,''} while P9 viewed the form more favourably, finding sections \textit{``more coherent''} and \textit{``a little bit more specific, especially with the evidence that they wanted.''}

\subsubsection{User Preferences} 
\begin{figure}[tb]
\raggedright
\includegraphics[width=1.05\columnwidth]{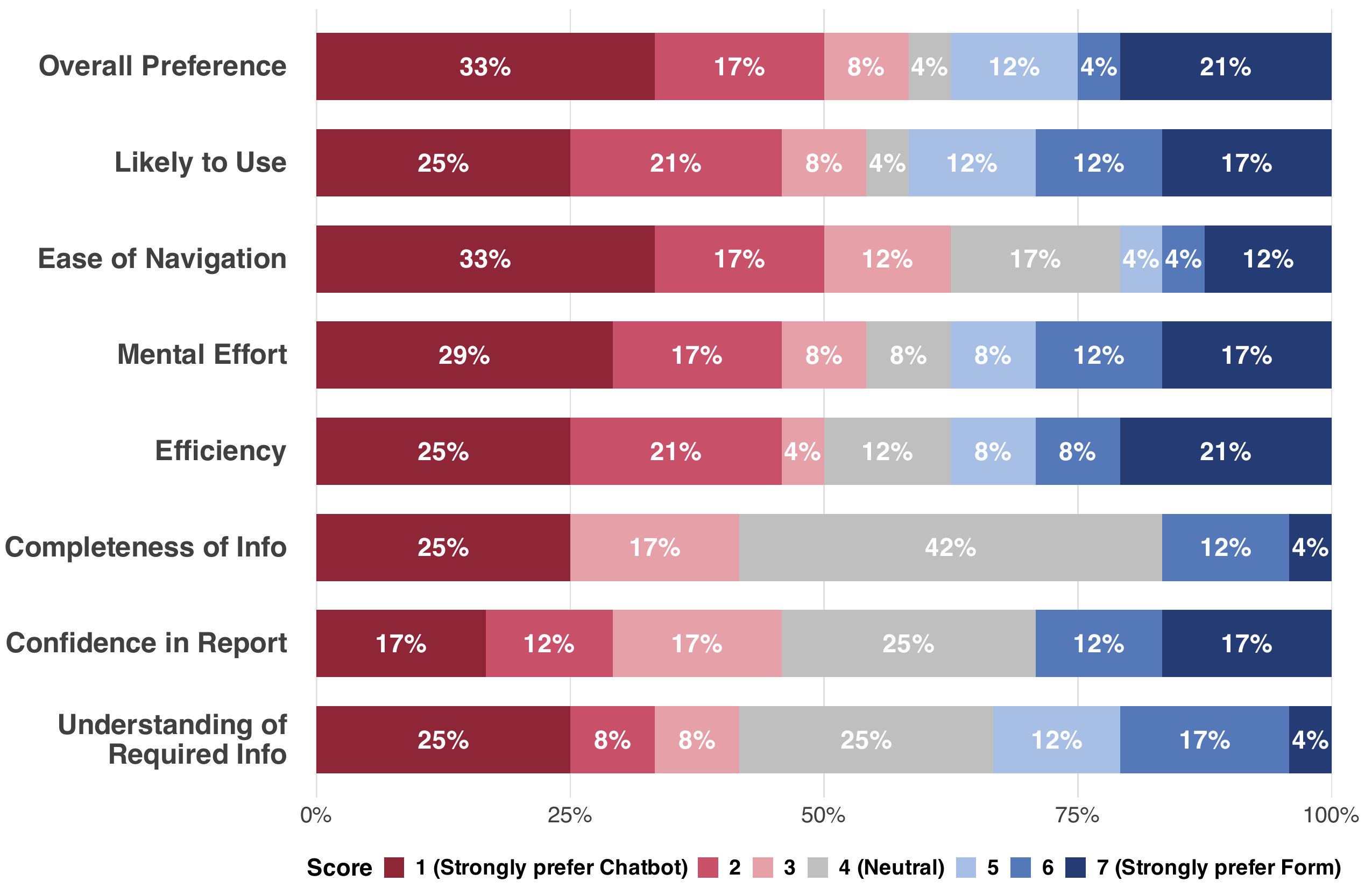}
\caption{Interface preferences by usability dimension}
\label{fig:better}
\end{figure}

Participants' comparative ratings show an overall preference for the chatbot across most usability dimensions (Figure~\ref{fig:better}). When asked about overall preference, 58\% preferred the chatbot, 38\% preferred the form, and 4\% were neutral ($\mu=3.42$, $\sigma=2.41$). The chatbot's clearest advantage emerged in navigation, where 63\% rated it as easier to navigate compared to 21\% for the form (18\% neutral, $\mu=3.04$, $\sigma=2.10$). Mental effort followed a similar trend, with 54\% of participants finding the chatbot less cognitively demanding, compared to 38\% favouring the form (8\% neutral, $\mu=3.54$, $\sigma=2.34$). Similarly, 50\% found the chatbot more efficient compared to 37\% for the form (12\% neutral, $\mu=3.67$, $\sigma=2.35$).

Preferences were more evenly distributed for understanding required information (42\% chatbot, 33\% form, 25\% neutral; $\mu=3.58$, $\sigma=1.95$) and confidence in the report (46\% chatbot, 29\% form, 25\% neutral; $\mu=3.83$, $\sigma=2.08$). Perceptions of information completeness were split, with 42\% favouring the chatbot, 17\% favouring the form, and 42\% remaining neutral ($\mu=3.46$, $\sigma=1.77$), suggesting that perceived completeness did not always align with objective report completeness. 

The remainder of this section summarizes key qualitative insights from this study. Although participants generally responded positively to the chatbot, we include a mix of positive and negative comments to highlight strengths as well as areas for improvement.  

\textbf{Flow and Guidance.} Participants frequently described the chatbot as more natural and supportive due to its conversational approach. P22 preferred that \textit{\enquote{the chatbot responded back to me, whereas the form didn't.}} P21 valued the chatbot's customization: \textit{\enquote{I quite enjoyed feeling like I was chatting with a tool that was customizing their responses to me}}. Similarly, P5 noted that \textit{\enquote{it was just easier to just talk and type out your thoughts and what's happening.}}

\textbf{Efficiency and Pacing.} While the chatbot’s guided structure was helpful for many, its sequential responses introduced issues with pacing. Some participants found that the conversational back-and-forth made the reporting process both slower and more cognitively demanding.
P11 explained, \textit{``It would take a bit longer compared to just inputting [information] in a form... you have to go bit by bit based on the questions.''} 
P17 was more blunt, stating that the chatbot \textit{\enquote{just kept going on [...] I just thought it was much more efficient if it just used a form instead.}} 
Others described pressure created by the back-and-forth interaction. 
P14 noted that because it felt like talking to a real person, they did not want to keep it waiting, which made the experience feel more pressured than using the form. P3 added that the need to wait for responses made the chatbot feel less efficient. Waiting for the chatbot’s responses frustrated participants who preferred to complete reports quickly. Part of this delay stemmed from variable AI response times, which may improve with newer models. However, even with faster systems, chatbots still present messages one at a time, unlike forms where users could view all questions at once.
P24, a self-described advanced user, found the chatbot \textit{\enquote{very, very cumbersome,}} explaining, \textit{\enquote{I can see it being really great for people that are maybe unfamiliar with things [...] because it helps guide them through, but for me, it's like, oh, come on, just give me a [form]. Let me type this stuff in.}}

\textbf{Reporting Challenges.} Participants encountered different challenges with each interface. With the form, vague or rigid wording often caused confusion. For the non-delivery scam, P14 struggled with the form question, \textit{``Has any merchandise been sold and or sent?''}, stating, \textit{``I don't understand what to answer for this,''} and selected an option that did not reflect their scenario. Similarly, P9 was unsure whether gift cards counted as \textit{``merchandise''} when reporting the CRA gift card scam. Participants also questioned the relevance of certain details. P16 chose not to include the scammer's name, explaining, \textit{``I don't even know if that's relevant because they're obviously using a fake name.''} These moments illustrate how uncertainty about terminology and the importance of fields can lead to incomplete or inaccurate reports.

\textbf{Trust.} As a prototype, the chatbot occasionally behaved unexpectedly. Trust in the chatbot was shaped by both malfunctions and prior experiences with AI. After learning that the chatbot could produce errors, P13 stated, \textit{``that would obviously make me more inclined to go with forms.''} Others were less concerned. P8 explained, \textit{``I've had [...] chatbots kind of messed up in the past as well... you always just have to double check''}.
Some participants linked reliability to institutional follow-through. P16 shared that malfunctions affected \textit{``how much I trust [that] the investigation will be taken seriously and done thoroughly.''} Even without errors, skepticism persisted for some. P15 admitted, \textit{``I guess part of me just doesn't [...] really trust chatbots''}.

The chatbot also raised trust concerns specific to the reporting context. When asked to share the giftcard number in the CRA scam call scenario, P23 hesitated, saying \textit{\enquote{I'm feeling a bit iffy about this. Why do I need to give you this information? I've just been scammed by giving a stranger a gift card number and details and I'm worried to do this again now.}} Although the chatbot's explanation about the information being used for investigations made sense to her, she ultimately chose to not provide that information. The initial reluctance highlights how the act of reporting can sometimes mirror the dynamics of the scam itself. In contrast, the form was often seen as more official and predictable. P9 explained, \textit{``With a form I knew what I wrote and there's only one way to interpret it.''} 
P16 said the form \textit{``felt much more official, like I could include all the information I wanted [...] in the manner I wanted.''} Trust in reporting interfaces is a complex topic; it is clearly important, but this trust can also be exploited by scammers launching similar-looking phishing sites against already vulnerable individuals.

\textbf{Emotional Tone.} Participants had mixed reactions to the chatbot's attempted empathy. For some, reassurance reduced stigma. P14 shared, \textit{``they reassured me... people who fall prey to scams, they feel dumb... it was empathetic.''} P13 reflected that it \textit{``kind of makes our sense of victimization feel like it's been heard, even though it hasn't, because it's a computer.''} P18 similarly appreciated \textit{\enquote{that human touch}} and \textit{\enquote{the lack of judgment}}. Others found empathetic messaging unnecessary or distracting. P12 stated, \textit{``I just think it's annoying... I don't need you, AI chatbots, sympathizing with me''}. 

The form's emotional tone was perceived as more procedural. P10 described it as \textit{``just, here, fill out this form, send it over, and then we'll look at it later.''} Still, some participants noticed subtle support elements. P9 appreciated \textit{``those little footnotes where it tells you like what you can do to prevent it.''} However, P18 perceived the form's repeated safety tips, such as warnings not to share personal information, as counterproductive. They described them as \textit{\enquote{a little, I guess, offensive,}} explaining that \textit{\enquote{the shame might hinder people and make them feel like it's their fault.}} Participants differed in whether they wanted emotional acknowledgement, P16 suggested adding \textit{``I'm sorry this happened to you,''} to the form, while P13 said they would likely ignore sympathy because they wanted to focus on completing the report.
\section{Discussion}

We discuss cybercrime reporting through four lenses: incident recognition, intention-action gap, trust in reporting systems, and the role of conversational interfaces, and conclude with recommendations and limitations.

\textbf{Recognition and Understanding.}
Participants' difficulty in recognizing their experience as cybercrime in our survey suggests that under-reporting may begin with a lack of awareness, not just practical barriers. This was also evident in the usability study, where participants often misinterpreted vague or technical wording in the online form, leading to incomplete or inaccurate reports. 
While prior work has mostly focused on structural or institutional barriers, such as perceptions that reporting will not lead to meaningful outcomes~\citep{bidgoli2016end}, confusion about where and how to report~\citep{ cross2016reporting}, and lack of confidence in police response~\citep{van2019determinants}, our findings highlight a more subtle but equally important set of internal, cognitive barriers, that occur earlier in the reporting process. Prior work indicates that reporting depends on whether victims perceive their experience as legitimate and worth reporting~\citep{graham2020willingness}, however, our findings show that some victims initially do not even identify their experience as cybercrime at all. This underscores a role for interface design in supporting earlier-stage sensemaking by helping users recognize, interpret, and assess the legitimacy of cyber incidents as they occur, rather than only supporting report submission.  
Designing reporting systems that address these early-stage cognitive barriers may lead to clearer, more complete reports and reduce drop-off due to confusion or uncertainty.

\textbf{Bridging Reporting Intentions and Action.}
Although perceived severity was positively associated with reporting likelihood, 
our findings suggest that translating reporting intentions into action may be influenced by additional individual factors beyond perceived harm. In our survey, women were more likely to report incidents, differing from prior findings~\citep{van2019determinants} that Dutch men report cybercrime more often and women are more likely to report traditional crimes. This may reflect evolving digital behaviours, cultural differences, or shifts in reporting patterns over time. 
In our study, technical skills were also a significant predictor, suggesting that tech-savvy individuals may feel more capable of navigating reporting processes or identifying reportable incidents, which in turn affects how victims act on their intentions. 
Together, these findings suggest that bridging the intention-action gap requires consideration of individual characteristics and the accessibility of reporting systems. This highlights the need for reporting tools that reduce procedural and cognitive challenges, adapt to user needs and capabilities, and effectively guide them through the overall reporting process.

\textbf{Trust as a Multi-layered Construct.}
Trust in cybercrime reporting operates across multiple interconnected layers. One established dimension is \textit{institutional trust}, or beliefs that authorities such as police or consumer protection agencies are fair, competent, and act in the public's best interest~\citep{spadaro2020enhancing}. In cybercrime contexts, this includes confidence that reports will be taken seriously and lead to meaningful action~\citep{cross2016reporting, jackson2025trust}. Our findings indicate that institutional trust alone is insufficient, as participants' reactions highlight the importance of \textit{technological trust}, \ie whether the reporting system is reliable and functional. When participants were told that the chatbot had skipped questions or failed to generate a summary, several questioned whether it could be trusted for real reporting, even within fictional scenarios, indicating sensitivity to small signs of failure. This aligns with prior work showing that cybercrime victims often become more wary of digital systems~\citep{cheng2020individual}. 
A third layer, \textit{interactional trust}, concerns whether the reporting process feels clear, respectful, and appropriate. 
Some participants struggled with the form's technical or ambiguous wording, while others found the chatbot's empathetic responses repetitive or inauthentic. 
Together, our results position trust in cybercrime reporting as layered and fragile; 
and it is built through small, cumulative moments rather than a single global judgment. 
Breakdowns at any one layer, whether institutional confidence, technological reliability, or interactional experience can all reduce users' willingness to engage. Reporting systems, therefore, should demonstrate dependable performance, communicate clearly and transparently, and use context-appropriate emotional cues to support user trust.

\textbf{AI within Technology-Mediated Reporting.}
AI chatbots are often described as encouraging open disclosure through their perceived non-judgmental nature~\citep{croes2024digital}. Consistent with this, participants generally preferred the chatbot over the form; however, not all participants valued the chatbot's conversational format. 

Some found it inefficient and its empathy artificial, aligning with findings by~\cite{bradford2025whom} showing that users preferred human operators over chatbots in crime reporting scenarios, even when responses were identical. This suggests that perceptions of trust and clarity are not just shaped by what is communicated, but by who or what is believed to be communicating it. These reservations, despite positive quantitative results, may also reflect a broader resistance to digital transformation \citep{scholkmann2021resistance, kherrazi2025managing}, where new systems require users to move beyond familiar reporting practices. 
Overall, our results indicate that the chatbot's strongest contribution was functional rather than relational. Participants valued its ability to prompt for missing details, reduce ambiguity, and provide a streamlined reporting flow, but any technical failures quickly reduced confidence in the system. 
This is especially relevant for individuals who already want to report but experience uncertainty about how to proceed, as the chatbot is most effective in supporting report completion by organizing information and clarifying the process rather than influencing report decisions.

\textbf{Recommendations.} 
Our findings point to several recommendations for policymakers, cybersecurity organizations, and product designers to strengthen cybercrime reporting systems and improve reporting rates. First, uncertainty about where to report was the most frequently cited reason for not reporting cybercrime across incident types. Relevant organizations should therefore collaborate to build a unified reporting platform to reduce confusion, streamline reporting, support consistent data collection, and improve responses across jurisdictions. Although such centralized systems exist in some jurisdictions, evaluations suggest they remain underused, indicating design gaps~\citep{morgan2016australia}. For example, a user-friendly browser extension that automatically capture relevant information (\eg suspicious URLs or phishing email content), could allow people to report cyber incidents immediately while details are fresh, saving time and improving report quality. Second, our usability study showed that chatbot glitches and inconsistent behaviour disrupted reporting flow and reduced participants' trust in the system to generate accurate reports. This highlights the importance of technical reliability, consistent performance, and transparent communication about data handling to maintain credibility and build user trust, especially for AI-driven systems where even minor errors can reduce confidence in the reporting process. Third, many participants struggled to recognize or correctly classify cyber incidents, creating barriers to reporting and contributing to incomplete or inaccurately categorized reports. Reporting systems should therefore incorporate simple recognition tools that guide users through clearly worded assessments using relatable scenarios to help users recognize a cybercrime and classify it. Finally, while cybercrime research and reporting practices often focus primarily on financial loss, our findings showed that time disruption was the strongest predictor of reporting behaviour, extending prior work on romance scam victims that highlighted emotional and psychological harms beyond monetary loss~\citep{whitty2016online}. Public awareness campaigns and reporting systems should therefore highlight the broader consequences of cybercrime, such as time theft, emotional disruption, and psychological harm. Shifting messaging from \textit{Did you lose money?} to \textit{Did this disrupt your peace of mind?}, and including framing like \textit{Help us stop criminals from targeting others} could broaden who recognizes their experience as worth reporting and acknowledges the emotional impact, while positioning reporting as a civic duty to protect the community and an empowering action, rather than requiring individuals to identify as victims.

\subsection{Study Limitations}
The online recruitment skewed our sample toward more technically literate participants than the general population. While concerns have been raised about data quality on crowdsourcing platforms such as Prolific, our study was conducted before widespread reports of automated or AI-assisted responses, and we did not observe such issues. Self-reported surveys introduce potential errors from misremembering, misclassifying past experiences, and social desirability biases~\citep{grimm2010social}. In the usability study, hypothetical scenarios cannot replicate the emotional intensity of actual victimization. Additionally, technical inconsistencies in the chatbot may have affected trust and report completeness. We suggest future work examine these factors and validate findings with victims in real-world reporting scenarios.

\section{Conclusion}

We examined cybercrime reporting behaviour through two complementary studies. 
Our survey identified key barriers to reporting, including difficulty recognizing victimization, and also revealed that time disruption 
had the strongest impact on reporting likelihood. Our usability study showed that AI chatbots better supported users by guiding them through the reporting process, reducing uncertainty, and producing more complete reports with higher 
satisfaction, but also identified 
improvement opportunities. Together, these studies provide insights into both why people report (or do not report) cybercrime and how interface design affects their reporting experience.

\section*{Acknowledgements}
We are grateful to our collaborators at the National Cybercrime Coordination Centre for their insight and feedback throughout this project. This work was funded through S. Chiasson's NSERC grants 571382-2022-SMFSU and RGPIN-2023-04653.

\bibliographystyle{agsm}
\balance
\bibliography{2026_ecrime_bib}
\appendix

\onecolumn
\section{Survey Questions}
\label{app.survey}

\textit{We had several repeated blocks of text. For simplicity, we label each block and use the corresponding label whenever the set is presented.}  

\textit{\textbf{Block A} consists of the following questions:}

\begingroup
\begin{itemize}[left=20pt,topsep=1pt,noitemsep,topsep=0pt,]
\item \textit{If yes:} Did you report this incident anywhere? [Yes, No]
\begin{itemize}[left=3pt,topsep=1pt,noitemsep,topsep=0pt,]
\item \textit{If yes:} Where did you report this incident? (Select all that apply) [Local police, Bank/financial institution, Platform where the incident occurred (for example, on a social media site, shopping website, email, or another online service), Canadian Anti-Fraud Centre 
Internet service provider,  
IT/Cybersecurity company,  
Consumer protection agency 
Other:]
\end{itemize}

\item \textit{If no:} Why didn't you report the incident? (Select all that apply) [Didn't think it was serious enough; Didn't know where to report it; 
Didn't think reporting would help; 
Felt embarrassed; 
Handled it myself; 
Too time consuming to report; 
Worried about privacy; 
Other:]
\end{itemize}

\textit{\textbf{Block B} consists of the following hypothetical scenario descriptions:}

\begin{itemize}[left=20pt,topsep=1pt,noitemsep,topsep=0pt,]
\item Your computer is not working as expected. You suspect that it has been infected with a virus and it needs to be fixed
\item Someone else gets into your online account without permission, and you can't log in anymore
\item You get alerts that someone tried to log into your email from another country
\item Due to a recent spike in your electricity bill, you suspect that your computer is being used to mine cryptocurrency without your permission
\item You lose \$300 in an online shopping scam and can't get your money back
\item You see someone online bragging about hacking a government website
\item You find fake job ads online trying to steal people's personal information
\item You see someone online selling other people's stolen personal information
\item You invest \$3,000 in an online opportunity that promises high returns. Later, you find out it's a scam, and you lose all your money
\item You receive threatening emails from a hacker who claims to have your personal photos
\item A hacker breaks into one of your social media accounts and sends embarrassing messages to all your contacts
\end{itemize}

\endgroup

\begin{enumerate}[left= 3pt,topsep=1pt,noitemsep]
    \item \textbf{Cybercrime:} Crimes where someone attacks computers or the internet, such as hacking into accounts, spreading viruses, or making websites and online services stop working. To your knowledge, have you ever fallen victim to a cybercrime incident as defined above? [Yes, No, I'm not sure]

\item To your knowledge, has anyone ever gained access to your online accounts without your permission? [Yes, No] \textbf{+ Block A}

\item To your knowledge, has your computer, phone, or tablet ever been infected with harmful software (like viruses or programs demanding payment to unlock your files)? [Yes, No] \textbf{+ Block A}

\item To your knowledge, have your personal details (such as name, social insurance number, email, password, address, credit card numbers, and other similar information) ever been exposed when a company's data was hacked? [Yes, No] \textbf{+ Block A}

\item Have you ever noticed any suspicious attempts to access your home network or personal devices? [Yes, No] \textbf{+ Block A}

\vspace{3pt}

\item \textbf{Cyber-enabled fraud:} Online scams used to steal money or personal information through fake websites, emails, or impersonation.   To your knowledge, have you ever fallen victim to a cyber-enabled fraud incident as defined above? [Yes, No, I'm not sure]

\item To your knowledge, have you ever lost money or personal information to an online scam (such as fake online stores, fake investment schemes, romance scams, job scams, and other scams)? [Yes, No] \textbf{+ Block A}

\vspace{3pt}

\item \textbf{Witnessing Cybercrime:} Seeing or hearing about online crimes happening to others, even if it didn't happen to you directly. For example, seeing scam posts on social media, knowing someone whose account was account was taken over without permission (hacking), or encountering suspicious content online.  Have you ever witnessed or become aware of a cybercrime incident not directly involving you?  [Yes, No, I'm not sure]

\item To your knowledge, have you ever seen posts online where people are selling stolen personal information or tools that could be used to illegally break into computers or online accounts? [Yes, No]~\textbf{+ Block A}

\item  To your knowledge, have you ever suspected that someone was using a computer or the internet for illegal activities? [Yes,~No] \textbf{+ Block A}

\item To your knowledge, have you ever encountered fake websites or emails trying to trick others into sharing their personal information? [Yes, No] \textbf{+ Block A}

\item To your knowledge, have you ever seen another person's social media account hacked or used by someone else without their permission? [Yes, No] \textbf{+ Block A}

\item To your knowledge, have you ever come across posts or discussions from people planning to hack or attack others' computers or accounts? [Yes, No] \textbf{+ Block A}

\item Of all the cybercrime and/or cyber-enabled fraud incidents you've experienced, which one had the biggest impact on you? 
[When someone attempted to gain unauthorized access to my network/devices,
When I lost money and/or personal information to an online scam, 
When someone gained unauthorized access to my online account(s), 
When my device was infected with harmful software, 
When my personal information was exposed in a data breach]
  \textbf{+ Block A}

\item Thinking about the incident you selected as having the biggest impact, how severe were the effects in these areas, based on your unique situation:  [No effect, Slightly severe, Moderately severe, Highly severe, Extremely severe]
 \begin{enumerate*}[label=(\roman*),left=5pt,topsep=1pt,noitemsep,topsep=0pt,]
\item Financial (cost to you or money lost)
\item Time (requires your time and effort to address)
\item Emotional (stress or worry caused by the incident)
\end{enumerate*}

\item How much do you agree or disagree with the following statements about the incident that greatly impacted you? [Strongly disagree, Disagree, Neither agree nor disagree, Agree, Strongly agree] 

 \begin{enumerate*}[label=(\roman*),left=5pt,topsep=1pt,noitemsep,topsep=0pt,]
\item I thought the incident was serious enough to report 
\item I decided to report the incident because of how it impacted me financially 
\item I decided to report the incident because of how much time it took to deal with it 
\item I decided to report the incident because of how it affected me emotionally
\item I didn't think the incident was important enough to report 
\item I wasn't sure if the incident was worth reporting 

\end{enumerate*}

\vspace{3pt}

\item The following are \textbf{hypothetical scenarios} involving cybercrime and cyber-enabled fraud. For each scenario, please indicate how likely you would be to \textbf{report} it: [Extremely unlikely, 
Somewhat unlikely, Neither likely nor unlikely, Somewhat likely, Extremely likely, I do not understand this scenario]  \textbf{+ Block B}

\item The following are hypothetical scenarios involving cybercrime and cyber-enabled fraud. For each scenario, please rate how \textbf{severe} you would consider this incident to be: [Not severe at all, Slightly severe, Moderately severe, Highly Severe, Extremely Severe, I do not understand this scenario]  \textbf{+ Block B}

\item The following are hypothetical scenarios involving cybercrime and cyber-enabled fraud. Based on your personal situation, categorize each scenario by what you think its \textbf{main impact} would be: [Financial (cost to you or money lost), Time (requires your time and effort to address), Emotional (stress or worry caused by the incident)] \textbf{+ Block B}

\vspace{3pt}

\item \textit{NGSE Scale~\citep{chen2001validation}}: Read each statement and choose how much you agree or disagree. There are no right or wrong answers. [Strongly disagree, Disagree, Neither agree nor disagree, Agree, Strongly agree] 

 \begin{enumerate*}[label=(\roman*),left=5pt,topsep=1pt,noitemsep,topsep=0pt,]
\item I will be able to achieve most of the goals that I set for myself
\item When facing difficult tasks, I am certain that I will accomplish them
\item In general, I think that I can obtain outcomes that are important to me
\item I believe I can succeed at most any endeavor to which I set my mind
\item I will be able to successfully overcome many challenges
\item I am confident that I can perform effectively on many different tasks
\item Compared to other people, I can do most tasks very well
\item Even when things are tough, I can perform quite well
\end{enumerate*}

\vspace{3pt}
\item \textit{AICS Scale~\citep{shulruf2007development}}: Please read the following questions and indicate how often you would think or behave as described in each of the following items.
[Never or almost never, Rarely, Occasionally, Often, Very often, Always] 

 \begin{enumerate*}[label=(\arabic*),left=5pt,topsep=1pt,noitemsep,topsep=0pt,]
\item I define myself as a competitive person
\item I enjoy being unique and different from others
\item Before I make a major decision I seek advice from people close to me
\item Even when I strongly disagree with my group members, I avoid an argument
\item I consult with superiors on work-related matters
\item I believe that competition is a law of nature
\item I prefer competitive rather than non-competitive recreational activities
\item Before taking a major trip, I consult with my friends
\item I sacrifice my self-interest for the benefit of my group
\item I consider my friends' opinions before taking important actions
\item I like to be accurate when I communicate
\item I consider myself as a unique person separate from others
\item It is important to consult close friends and get their ideas before making a decision
\item Without competition, I believe, it is not possible to have a good society
\item I ask the advice of my friends before making career-related decisions
\item I prefer using indirect language rather than upsetting my friends by telling them directly what they may not like to hear
\item It is important for me to act as an independent person
\item I discuss job or study-related problems with my parents/partner
\item I take responsibility for my own actions
\item I do not reveal my thoughts when it might initiate a dispute
\item I try to achieve better grades than my peers
\item My personal identity independent of others is very important to me
\item I consult my family before making an important decision
\item Winning is very important to me
\item I see myself as ``my own person''
\item I enjoy working in situations involving competition with others
\end{enumerate*}

\vspace{3pt}

\item How would you rate your overall computer/internet skills?   [Very advanced: I can troubleshoot most problems and often help others, Advanced: I can solve common technical problems myself, 
Intermediate: I can do most everyday tasks but might need help, Basic: I can do simple tasks like email and web browsing but often need help, Beginner: I am just learning to use computers/the internet]

\item How comfortable would you be reporting a cybercrime in English? 
  [Very comfortable: I can explain technical problems clearly, Comfortable: I can explain most things with some effort, 
Somewhat comfortable: I might need help with certain terms, Uncomfortable: I would have trouble explaining the problem, Very uncomfortable: I would need someone to translate]

\item How comfortable would you be reporting a cybercrime in French? 
  [Very comfortable: I can explain technical problems clearly, Comfortable: I can explain most things with some effort, 
Somewhat comfortable: I might need help with certain terms, Uncomfortable: I would have trouble explaining the problem, Very uncomfortable: I would need someone to translate] 

\item Would it help if you could report in another language? [Yes, No, Not Sure]

\item How old are you? 
\item What is your gender?   [Man, Woman, Non-binary / third gender, Prefer not to say,
 Not listed: please describe]
 
\item What is the highest level of education you have completed?   [Less than high school 
High school graduate, Some university/college, 2-year diploma, 3-year diploma, 4-year degree, Graduate or professional degree, Doctorate, Other]

\item How do you identify your race/ethnicity? (select all that apply)   
[African/Black (including African-American, African-Canadian, Caribbean);  
East Asian (Chinese, Taiwanese, Japanese, Korean, etc.);  
European/White;
Indo-Caribbean, Indo-African, Indo-Fijian, West-Indian;  
Latin, South or Central American;  
Polynesian (Samoans, Tongan, Niuean, Cook Island Maori, Tahitian, Hawaiian, Marquesan, New Zealand Maori);  
South Asian (Afghan, Nepali, Tamil, Bangladeshi, Pakistani, Indian, Sri Lankan, Punjabi);  
Southeast Asian (Vietnamese, Thai, Cambodian, Malaysian, Filipino/a, Laotian, Singaporean, Indonesian);
West Asian (Iraqi, Jordanian, Palestinian, Saudi, Syrian, Yemeni, Armenian, Iranian, Israeli, Turkish);  
Indigenous within Canada (First Nation, Métis, Inuit);  
Prefer to self-identify; Prefer not to answer]

\item Including yourself, how many people live in your household?

\item What was your total household income before taxes during the past 12 months?
[Less than \$55,867; \$55,867 to \$111,733;  \$111,733 to \$173,205;  \$173,205 to \$246,752; Above \$246,752]

\end{enumerate} 

\section{Usability Study Materials}
\label{app.usability}
\subsection{Usability Study Scenarios}

 \textbf{Verbal Instructions.} You’ll be presented with two different cyber incident scenarios.
For each scenario: Take a moment to read and understand the situation. Use the provided tool to report the incident. Share your thoughts out loud as you work through the process.
Remember to:

\begin{itemize}[left= 3pt,topsep=1pt,noitemsep]
  \item Make it your own: the scenarios provide a basic outline, but feel free to fill in details that make sense to you. There’s no right or wrong way to interpret them.
  \item Report as you naturally would: imagine this actually happened to you. How much information would you share? What details would seem important to include? What would you leave out?
  \item Think out loud: as you work through the reporting process, please share your thoughts, frustrations, questions, and decisions. Say things like:
“I’m looking for where to enter the date.”, “I’m not sure what they mean by this question.”,  “I think this part is important to include because…”
  \item There are no right or wrong answers: we’re testing the tools, not you. Your natural response is exactly what we want to see.
\end{itemize}

\begin{figure}[h]
  \centering
  \begin{minipage}[t]{0.4\textwidth}
    \centering
    \includegraphics[width=.7\linewidth]{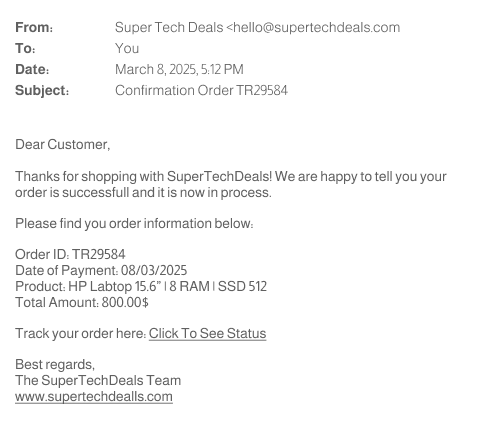}
    \subcaption{Order confirmation email}
  \end{minipage}
  \begin{minipage}[t]{0.4\textwidth}
    \centering
    \includegraphics[width=.7\linewidth]{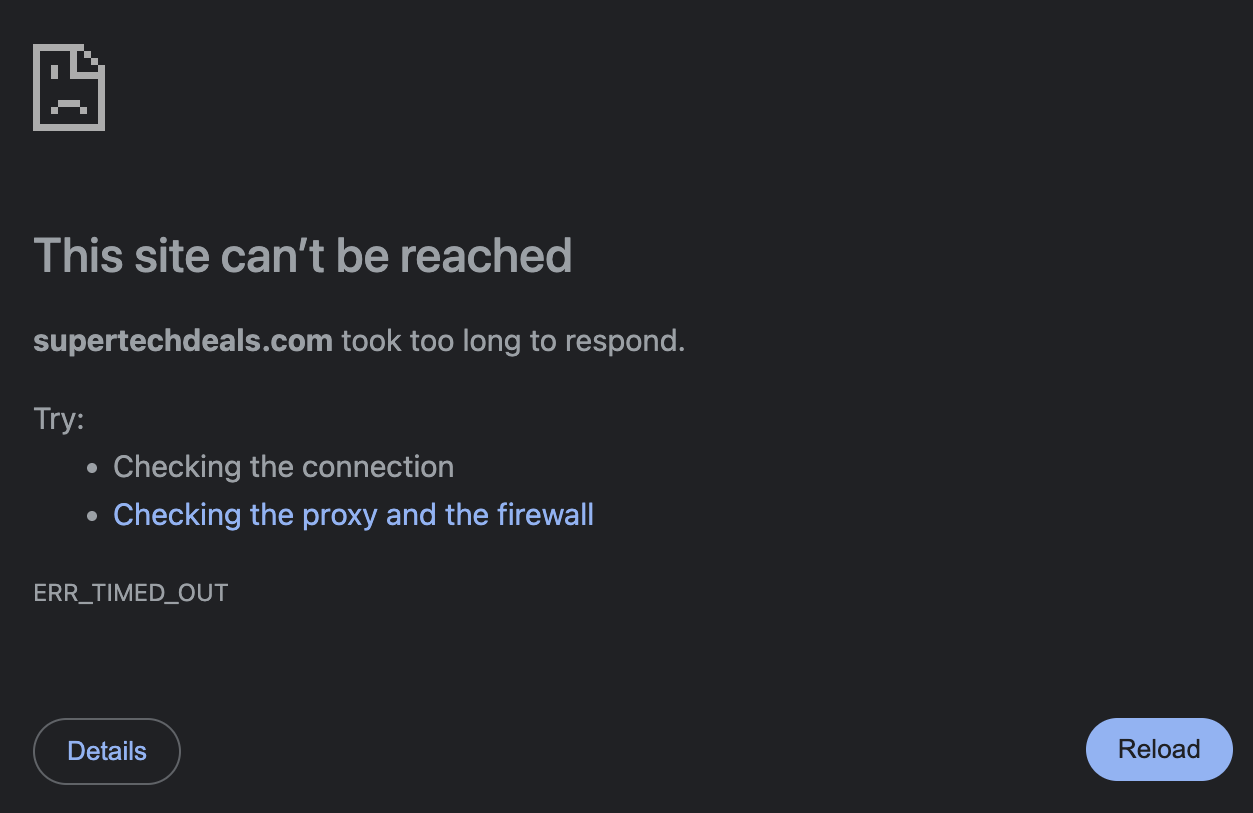}
    \subcaption{Screenshot showing the original website is down}
  \end{minipage}
  \caption{Non-Delivery Scam: Optional evidence to include in non-delivery scam report}
  \label{scam_nondelivery}
\end{figure}

\textbf{Non-Delivery Scam.}
\textit{Participants were given the following information for the Non-Delivery scam, along with the two images found in Figure~\ref{scam_nondelivery}.}   

You bought a laptop online for \$800. The store had good reviews and seemed real. After paying with your credit card, you got an email with your order number and a tracking link. Two weeks have passed, and:
    \begin{enumerate*}[label=(\roman*),left= 3pt,topsep=1pt,noitemsep]
        \item Your package hasn't arrived
        \item The tracking number doesn't show any updates
        \item Nobody answers your emails to customer service
        \item The store's phone number doesn't work anymore
        \item The website is now completely gone
        \item The \$800 charge is still on your credit card
    \end{enumerate*}
    
You want to report this. When you report, remember these details:
     \begin{enumerate*}[label=(\roman*),left= 3pt,topsep=1pt,noitemsep]
        \item Order \#TR29584
        \item Date of purchase: March 8, 2025
        \item Payment method: Credit card
        \item Last email sent to company: March 18, 2025
        \item Website name: \texttt{SuperTechDeals.com}
    \end{enumerate*}

\begin{figure}[h]
  \centering
  \begin{minipage}[t]{0.4\textwidth}
    \centering
    \includegraphics[width=.5\linewidth]{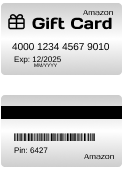}
    \subcaption{Gift card}
  \end{minipage}
  \begin{minipage}[t]{0.4\textwidth}
    \centering
    \includegraphics[width=.6\linewidth]{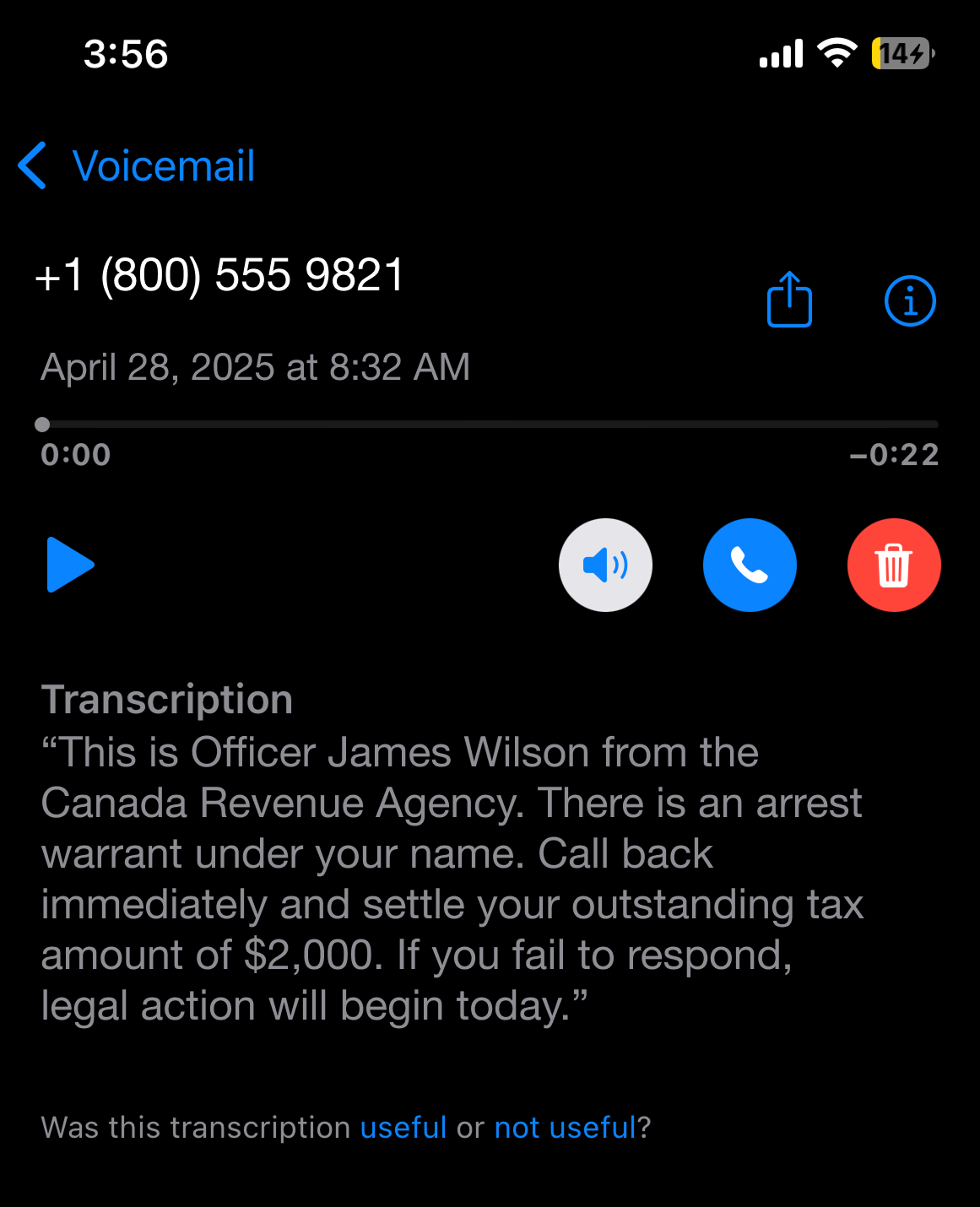}
    \subcaption{Voicemail transcription}
  \end{minipage}
  \caption{Canada Revenue Agency Scam: Optional evidence to include in CRA scam report}
  \label{scam_cra}
\end{figure}

\textbf{Canada Revenue Agency Scam.}
\textit{Participants were given the following information for the Canada Revenue Agency scam, along with the two images found in Figure~\ref{scam_cra}.}   

You got a phone call from someone claiming to be from the Canada Revenue Agency (CRA). They said you owed \$2,000 in taxes and would be arrested if you didn’t pay immediately by gift card.
They knew your name and address, and you were scared, so you followed their instructions to:
     \begin{enumerate*}[label=(\roman*),left= 3pt,topsep=1pt,noitemsep]
        \item Buy a \$2,000 Amazon gift card
        \item Scratch off the code and read it over the phone
        \item After giving them the code, they hung up
        \item You later realized the CRA doesn't collect taxes by gift card
    \end{enumerate*}
    
You want to report this. When you report, remember these details:
     \begin{enumerate*}[label=(\roman*),left= 3pt,topsep=1pt,noitemsep]
        \item Date of call: April 28, 2025
        \item Caller's phone number: 1-800-555-9821
        \item Caller's name: Officer James Wilson
        \item Gift Card Number: 4000 1234 4567 9010
        \item Gift Card Activation Number: 6427
    \end{enumerate*}

\subsection{Usability Study Questionnaires}
\label{app.usability_questionnaires}

\textbf{Post-Task Questions}
\textit{These questions were completed immediately after each scenario.}

\begin{enumerate}[left= 3pt,topsep=1pt,noitemsep]
    \item How satisfied were you with this reporting tool?    [1 = Very dissatisfied, 7 = Very satisfied]

    \item How difficult or easy was reporting with this tool?    [1 = Very difficult, 7 = Very easy]
\end{enumerate}

\textbf{Post-Session Questionnaire} \textit{These questions were completed after both tasks were done.}

\begin{enumerate}[left= 3pt,topsep=1pt,noitemsep]
\item Which reporting tool did you prefer overall?   [1 = Strongly prefer chatbot, 7 = Strongly prefer form]

\item Which tool would you be more likely to use if you experienced an actual cyber incident? 
  [1 = Much more likely to use chatbot, 7 = Much more likely to use form]

\item Which tool was easier to navigate? 
  [1 = Chatbot much easier, 7 = Form much easier]

\item Which tool required less mental effort to complete?   [1 = Chatbot required much less effort, 7 = Form required much less effort]

\item Which tool felt more efficient?   [1 = Chatbot much more efficient, 7 = Form much more efficient]

\item Which tool helped you provide more complete information?   [1 = Chatbot helped provide much more complete information, 7 = Form helped provide much more complete information]

\item Which tool made you feel more confident your report would be properly addressed?   [1 = Much more confident with chatbot, 7 = Much more confident with form]

\item Which tool provided a better understanding of what information was needed?   [1 = Chatbot provided much better understanding, 7 = Form provided much better understanding]
\end{enumerate}

\textbf{Experience Questions}
\begin{enumerate}[left= 3pt,topsep=1pt,noitemsep]
\item How likely would you be to report a cyber incident generally?   [1 = Not at all likely,7 = Very likely]
\item How likely would you be to report a cyber incident using the form?   [1 = Not at all likely, 7 = Very likely]
\item How likely would you be to report a cyber incident using the chatbot?   [1 = Not at all likely, 7 = Very likely]
\end{enumerate}

\textbf{Demographic Questions}
\begin{enumerate}[left= 3pt,topsep=1pt,noitemsep]

\item How would you rate your overall computer/internet skills?   [Very advanced: I can troubleshoot most problems and often help others, Advanced: I can solve common technical problems myself, 
Intermediate: I can do most everyday tasks but might need help, Basic: I can do simple tasks like email and web browsing but often need help, Beginner: I am just learning to use computers/the internet]

\item How comfortable would you be reporting a cybercrime in English? 
  [Very comfortable: I can explain technical problems clearly, Comfortable: I can explain most things with some effort, 
Somewhat comfortable: I might need help with certain terms, Uncomfortable: I would have trouble explaining the problem, Very uncomfortable: I would need someone to translate] 

\item How old are you? 
\item What is your gender?   [Man, Woman, Non-binary / third gender, Prefer not to say,
 Not listed: please describe]
 
\item What is the highest level of education you have completed?   [Less than high school 
High school graduate, Some university/college, 2-year diploma, 3-year diploma, 4-year degree, Graduate or professional degree, Doctorate, Other]

\item How do you identify your race/ethnicity? (select all that apply)   
[African/Black (including African-American, African-Canadian, Caribbean);  
East Asian (Chinese, Taiwanese, Japanese, Korean, etc.);  
European/White;
Indo-Caribbean, Indo-African, Indo-Fijian, West-Indian;  
Latin, South or Central American;  
Polynesian (Samoans, Tongan, Niuean, Cook Island Maori, Tahitian, Hawaiian, Marquesan, New Zealand Maori);  
South Asian (Afghan, Nepali, Tamil, Bangladeshi, Pakistani, Indian, Sri Lankan, Punjabi);  
Southeast Asian (Vietnamese, Thai, Cambodian, Malaysian, Filipino/a, Laotian, Singaporean, Indonesian);
West Asian (Iraqi, Jordanian, Palestinian, Saudi, Syrian, Yemeni, Armenian, Iranian, Israeli, Turkish);  
Indigenous within Canada (First Nation, Métis, Inuit);  
Prefer to self-identify; Prefer not to answer]   
\end{enumerate}

\subsection{Usability Study Interview Questions}
\label{app.usability_interview}

\textbf{Background Questions on Cyber Incident Reporting}
Have you ever experienced a cyber incident? A cyber incident includes any situation where someone uses the internet or technology to harm, deceive, or steal from you (for example, phishing, online scams, identity theft, hacking accounts, malware attacks, and unauthorized access to your information or devices).

\textit{(If yes)} Did you report the incident? Who did you report it to (e.g., a bank, police, fraud center, tech support)? Can you describe your experience reporting it? Were there any challenges or difficulties in the process?

\textit{(If no)} Why did you decide not to report it? Were you unsure where to report it? Did you feel it wasn’t serious enough? Did you think nothing would be done about it?

\textbf{Post-session interview questions}
\begin{enumerate*}[label=(\roman*),left= 3pt,topsep=1pt,noitemsep]
  \item What did you like most about the form? The chatbot?
  \item What challenges did you encounter with each tool?
  \item What would make you more likely to report a cyber incident in the future?
  \item Is there any information you felt was difficult to provide in either tool?
  \item Did either tool prompt you to include details you might not have thought to mention?
  \item Is there anything else you would like me to know?
\end{enumerate*}

\end{document}